\documentclass[11pt,a4paper]{article}
\pdfoutput=1
\usepackage{jcappub}

\usepackage{graphicx,amsmath,amsfonts,amssymb,revsymb,dcolumn,bm,mathrsfs,enumerate,epsfig}

\usepackage{pdflscape}
\usepackage{amsmath}
\usepackage{amssymb}
\usepackage{dcolumn}
\usepackage{bm}
\usepackage{color}
\usepackage{epsfig}
\usepackage{amsfonts}
\usepackage{graphicx}
\usepackage{subfigure}
\usepackage{dcolumn}

\newcommand{\be}{\begin{equation}}
\newcommand{\ee}{\end{equation}}
\newcommand{\bea}{\begin{eqnarray}}
\newcommand{\eea}{\end{eqnarray}}

\usepackage{amsmath, amssymb, graphics, setspace}
\newcommand{\mathsym}[1]{{}}
\newcommand{\unicode}[1]{{}}
\usepackage{textcomp}
\usepackage{eurosym}
\usepackage{amsfonts}
\usepackage{array}
\usepackage{amsthm}
\usepackage{bm}
\usepackage{palatino}

\usepackage{hyperref}

\begin{document}

\title{Inflation from a nonlinear magnetic monopole field nonminimally coupled to curvature}

\author[a]{Giovanni Otalora}
\author[a,b,c,d]{Ali \"{O}vg\"{u}n}
\author[a]{Joel Saavedra}
\author[a]{Nelson Videla}

\affiliation[a]{Instituto de F\'{\i}sica, Pontificia Universidad Cat\'olica de
Valpara\'{\i}so\\ Casilla 4950, Valpara\'{\i}so, Chile}

\affiliation[b]{Physics Department, Arts and Sciences Faculty, Eastern Mediterranean University, Famagusta, North Cyprus via Mersin 10, Turkey}

\affiliation[c]{Physics Department, California State University Fresno, Fresno, CA 93740,USA}

\affiliation[d]{Stanford Institute for Theoretical Physics, Stanford University, Stanford,
CA 94305-4060, USA}

\emailAdd{giovanni.otalora@pucv.cl}
\emailAdd{ali.ovgun@pucv.cl}
\emailAdd{joel.saavedra@pucv.cl}
\emailAdd{nelson.videla@pucv.cl}

\abstract{
In the context of nonminimally coupled $f(R)$ gravity theories, we study early inflation driven by a nonlinear monopole magnetic field which is nonminimally coupled to curvature. In order to isolate the effects of the nonminimal coupling between matter and curvature we assume the pure gravitational sector to have the Einstein-Hilbert form. Thus, we study the most simple model with a nonminimal coupling function which is linear in the Ricci scalar. From an effective fluid description, we show the existence of an early exponential expansion regime of the Universe, followed by a transition to a radiation-dominated era. In particular, by applying the most recent results of the Planck collaboration we set the limits on the parameter of the nonminimal coupling, and the quotient of the nonminimal coupling and the nonlinear monopole magnetic scales. We found that these parameters must take large values in order to satisfy the observational constraints. Furthermore, by obtaining the relation for the graviton mass, we show the consistency of our results with the recent gravitational wave data GW$170817$ of LIGO and Virgo. }


\maketitle

\section{Introduction}
\label{Introduction}

Cosmic inflation is the currently accepted paradigm which provides an elegant solution to the well-know problems of standard Big Bang cosmology \cite{Guth:1980zm,Linde:1981mu,Starobinsky}. Furthermore, it allows us to explain the origin of primordial fluctuations, which are the seeds for all observed structures in the Universe \cite{Mukhanov:1990me,Liddle_Lyth2000}. The early inflationary phase can be studied in the context of scalar field models by maintaining General Relativity (GR) as the gravitational theory, where a canonical scalar field $\phi$ (inflaton) drives a quasi-exponential expansion through an arbitrary flat potential $V(\phi)$ \cite{MukhanovBook, Liddle:1999mq}. Another approach is obtained in the context of modified gravity models (e.g. $f(R)$ gravity theories \cite{DeFelice:2010aj, Sotiriou:2008rp, Nojiri:2010wj, Lobo:2008sg, Clifton:2011jh, Capozziello:2011et,Copeland:2006wr, Nojiri:2017ncd}), where higher-order curvatures terms, motivated from quantum corrections, are also added to the Einstein-Hilbert action of GR. 


The most general scenario, which mixes the best attributes and potentialities of each one of the above approaches, is obtained by allowing a nonminimal coupling between curvature and matter \cite{Nojiri:2004bi,Allemandi:2005qs,Nojiri:2006ri,Bertolami:2007gv,Harko:2008qz,Harko:2010mv}. In these nonminimal matter-curvature coupling theories, besides of an arbitrary function $f_{1}(R)$ in the gravity sector, one also introduces a nonminimal coupling by adding to the action a term of the form $f_{2}(R) \mathcal{L}_{m}$, where $f_{2}(R)$ is an arbitrary function of the scalar curvature and $\mathcal{L}_{m}$ is the matter Lagrangian density. It has been shown in Ref. \cite{Bertolami:2009ic} that these theories allow to mimic known dark matter density profiles through an appropriate power-law coupling term with negative power index, in such a way that it is guaranteed the dominance of dark matter, before the dark energy-dominated era. A study of the dynamics of cosmological perturbations in these theories has been performed in Ref. \cite{Bertolami:2013kca}, where the authors have examined how the evolution of cosmological perturbations is affected due to the presence of a nonminimal coupling between curvature and matter on a homogeneous and isotropic Universe, and hence the formation of large-scale structure. The late-time acceleration of the Universe was studied in Refs. \cite{Bertolami:2011fz, Bertolami:2010cw}, where it has been shown that the current cosmic acceleration may have originated from a nonminimal curvature-matter coupling, and, through mimicking, it can provide a viable scenario for the unification of dark energy and dark matter. Moreover, recently in Ref. \cite{Gomes:2016cwj}, the authors have investigated an inflationary scenario in this context, by taking the matter Lagrangian density to be that one of a canonical scalar field. Thus, they have shown that the spectrum of primordial perturbations and the inflationary dynamics can be significantly modified once that the energy density of matter  exceeds a critical value which is determined by the scale of the nonminimal coupling term. Finally, the detection of gravitational waves with an electromagnetic counterpart (GW$170817$) \cite{TheLIGOScientific:2017qsa}, emitted by a binary neutron star merger, has put a strong constraint on the speed of tensor polarization modes, which becomes extremely very close to the speed of light with a deviation smaller than $10^{-15}$ \cite{Lombriser:2015sxa,Lombriser:2016yzn,Monitor:2017mdv,Baker:2017hug,Ezquiaga:2017ekz,Creminelli:2017sry,Sakstein:2017xjx}. So, in Ref. \cite{Bertolami:2017svl}, it was also recently analysed the propagation of gravitational waves modes in nonminimal curvature-matter theories, for the cases of cosmological constant and dark energy-like fluid as matter source, showing a good agreement with the recent gravitational wave constraints.

Nevertheless, in choosing the matter Lagrangian for the very early Universe, the scalar field Lagrangian is not the only possibility. 
Big Bang, which was the starting point of the Universe, has a singularity \cite{haw}, and this singularity breaks all the physics laws \cite{Carrol}, therefore being this the big gap in the understanding of nature of the Universe \cite{MukhanovBook}. Nowadays, the various studies on nonlinear electromagnetic fields are done to remove the singularity of the Universe \cite{Durrer:2013pga,Kunze:2013kza,Kunze:2007ph}. The main idea is to modify the Maxwell equations smartly to remove singularity \cite{Novello:2003kh}. Using this fundamental method at strong fields such as early universe, nonsingular universe models have been found, namely magnetic universe \cite{novello0}, however sometimes it is difficult to preserve the conformal invariance of Maxwell fields after the modification of the electromagnetic fields. Magnetic fields are believed to have played a large part in the dynamics of the evolution of our universe. However, little is known about the existence of magnetic fields when the universe was very young. Observations have manifested the existence of magnetic field in the universe, ranging from the stellar scale $10^{-5}$ pc to the cosmological scale $10^4$ $Mpc$ \cite{Neronov:1900zz,Taylor:2011bn}. In particular, the magnetic field on large scales ($\leq 1$ $Mpc$) is deemed to be produced in the early universe \cite{Subramanian:2015lua}, namely, the primordial magnetic field. By the recent CMB observations, the strength of the magnetic field is smaller than a few nano-gauss at the $1$ $Mpc$ scale \cite{Ade:2015cva}. Additionally, the $\gamma-ray$ detections of the distant blazars imply that the magnetic field should be larger than $10^{-16}$ gauss on the scales $1 - 10^4$ $Mpc$ \cite{AlvesBatista:2016urk}.

Although there are no direct observations of primordial magnetic fields, it is believed that they existed because may have been needed to seed the large magnetic fields observed today. Theories also disagree on the amplitude of primordial magnetic fields. There are currently several dozen of theories about the origin of cosmic magnetic fields for instance primordial vorticity plasma (vortical motion during the radiation era of the early Universe, vortical thermal background by macroscopic parity-violating currents), quantum-chromo-dynamics phase transition, first-order electroweak phase transition via a dynamo mechanism, etc (\cite{Grasso:2000wj} and references therein). In the last scenario, seed fields are provided by random magnetic field
fluctuations which are always present on a scale of the order of a thermal wavelength. It refers to stochastic background with a nonzero value of $< B^2 >$. Stochastic magnetic fields are the cosmic background with the wavelength smaller than the curvature. Thermal fluctuations in a dissipative plasma, i.e. plasma fluctuations, could source  stochastic magnetic fields on a scale larger than the thermal wavelength \cite{Grasso:2000wj}. Thus, there are the stochastic fluctuations of the electromagnetic field in a relativistic electron–positron plasma. Although homogeneous magnetic fields can affect  the isotropy of the Universe, i.e. the energy-momentum tensor can become anisotropic which could cause an anisotropic expansion law and modify the CMB spectrum, the effect on the Universe geometry (isotropy) of magnetic fields  tangled on scales much smaller than the Hubble radius are negligible \cite{Grasso:2000wj}. Thus, averaging the magnetic fields, which are sources in general relativity \cite{tolman}, give the isotropy of the Friedman-Robertson-Walker (FRW) spacetime. Symmetry arguments also can be used to construct a  triplet of mutually orthogonal (magnetic) vector fields which respects in an exact form the spatial homogeneity and isotropy of the Universe \cite{Bento:1992wy,Bertolami:1991cf,Bertolami:1990je}. Although, $N$ randomly oriented vector fields lead to a slightly anisotropic Universe with a degree of anisotropy of order $1/\sqrt{N}$ at the end of inflation \cite{Golovnev:2008cf}, one can avoid a large anisotropy by considering a large number of randomly oriented vector fields in a sufficiently large time-dependent three-volume, as through of the average procedure used in Refs. \cite{DeLorenci:2002mi,Novello:2003kh}. So, as it has been shown in Refs. \cite{Novello:2003kh, Novello:2006ng, novello0, Ovgun:2016oit,Ovgun:2017iwg, Kruglov:2017vca, Sharif:2017pdd, Kruglov:2014hpa, Kruglov:2016cdm, Kruglov:2015fbl, Kruglov:2016lqd,Campanelli:2007cg, Hollenstein:2008hp, Bamba:2008ja, Durrer:2013pga, Kunze:2013kza, Kunze:2007ph}, a nonlinear magnetic monopole field can also play an important role in the inflationary dynamics, despite breaking the conformal invariance of the Maxwell theory \cite{Kunze:2013kza, Kunze:2007ph}. Furthermore, since a nonminimal coupling with curvature can affect significantly the physical results obtained in these models \cite{Gomes:2016cwj}, here, we are going to study a nonlinear magnetic monopole field nonminimally coupled to curvature and its implications for the early inflationary phase.

The paper is organized as follows. In Section II, we define our model on the nonminimal matter-curvature theories and in Section III, we study the effect of nonlinear electrodynamics on the cosmology using the nonminimal matter-curvature coupling. In Section IV, the evolution of the Universe is studied and we introduce an effective perfect fluid description, which allows us to put the modified Friedmann equations in its effective standard form, by identifying the relevant cosmological parameters. Thus, in order to study the dynamics of inflation, we introduce the slow-roll parameters, along with the spectral indices to comparison with observational data. 
Finally, in Section V, we conclude our paper.

\section{Nonminimal matter-curvature theories}\label{CovF(T)}
Our starting point is the action principle \cite{Bertolami:2007gv}:
\be
S=\int{d^{4}x\sqrt{-g}\left[\frac{1}{2 \kappa^2}f_{1}(R)+f_{2}(R) \mathcal{L}_{m}\right]},
\label{Nf(R)}
\ee where $\kappa^2=8 \pi G$, $f_{1}(R)$ and $f_{2}(R)$ are functions of the Ricci scalar, and $\mathcal{L}_{m}$ is the matter Lagrangian density.

Varying the action \eqref{Nf(R)} with respect to the metric we obtain the field equations
\be
\left(R_{\mu\nu}-\Delta_{\mu\nu}\right)\left[F_{1}+2 \kappa^2 \mathcal{L}_{m} F_{2} \right]-\frac{1}{2} f_{1} g_{\mu\nu}=\kappa^2 f_{2} T_{\mu\nu}^{(m)},
\label{FEq}
\ee
where we have defined $F_{i}(R)=f'_{i}(R)$, being that prime represents the derivative with respect to the Ricci scalar, and $\Delta_{\mu\nu}=\left(\nabla_{\mu}\nabla_{\nu}-g_{\mu\nu}\Box\right)$.The matter energy-momentum tensor is defined as
\be
T_{\mu\nu}^{m}=-\frac{2}{\sqrt{-g}}\frac{\delta{\left(\sqrt{-g}\mathcal{L}_{m}\right)}}{\delta g^{\mu\nu}}.
\label{EMT}
\ee From the field equations \eqref{FEq} and by using the Bianchi identities we can show that the matter energy-momentum tensor satisfies the conservation law
\be
\nabla^{\mu} T_{\mu\nu}^{m}=\frac{F_{2}}{f_{2}}\left[g_{\mu\nu}\mathcal{L}_{m}-
 T_{\mu\nu}^{m}\right]\nabla^{\mu} R,
\label{EClaw}
\ee which describes an exchange of energy and momentum between matter and the higher
derivative curvature terms \cite{Bertolami:2007gv}. In the next section we will assume that the matter Lagrangian $\mathcal{L}_{m}=\mathcal{L}_{NMM}$ is the Lagrangian of a nonlinear magnetic monopole field. 

\section{Non-linear magnetic monopole fields
and cosmology}

The Lagrangian density of the non-linear magnetic monopole is given by \cite{Kruglov:2014hpa}
\be
\mathcal{L}_{m}=\mathcal{L}_{NMM}=-\frac{\mathcal{F}}{1+2\beta \mathcal{F}},
\label{NMMF}
\ee where $\mathcal{F}$ is the Maxwell invariant. Since the matter Lagrangian does not depend on the derivatives of the metric, and using the definition \eqref{EMT}, the matter energy-momentum tensor is written as \cite{Kruglov:2017vca}
\be
T_{\mu\nu}=-K_{\mu \lambda} F_{\nu}^{~\lambda}+g_{\mu\nu} \mathcal{L}_{NMM},
\label{EMT1}
\ee 
with
\be
K_{\mu \lambda}=\frac{\partial \mathcal{L}_{NMM}}{\partial \mathcal{F}} F_{\mu \lambda}.
\ee
We suppose that there is a dominant stochastic
magnetic fields on the cosmic background and their wavelengths are smaller than the curvature so that the averaging of EM fields are used as a source of Einstein equations coupled to non-minimal coupled to curvature
\cite{tolman}. The averaged electromagnetic fields are given as \cite{DeLorenci:2002mi,Novello:2003kh}:

\begin{equation}
\langle E\rangle=\langle B\rangle=0,\text{ }\langle E_{i}B_{j}\rangle=0,
\end{equation}
\[
\langle E_{i}E_{j}\rangle=\frac{1}{3}E^{2}g_{ij},\text{ }\langle B_{i}B_{j}\rangle=\frac{1}{3}B^{2}g_{ij},
\]
which $\langle$ $\rangle$ is the averaging brackets. We consider it to take average volume and this volume is larger value than the radiation wavelength, and also smaller than the curvature.

Thus, in the pure non-linear magnetic monopole case, such that $E^2= 0$,
from Eq. \eqref{EMT1} we obtain the energy density $\rho=-T^{0}_{~0}$ and the pressure $p=T^{i}_{~i}/3$ of the non-linear monopole magnetic field \cite{Ovgun:2016oit}
\bea
\label{rho}
&&\rho=-\mathcal{L}_{NMM},\\
&& p=\mathcal{L}_{NMM}-\frac{2 B^2}{3}\frac{\partial \mathcal{L}_{NMM}}{\partial \mathcal{F}}, 
\label{p}
\eea where $\mathcal{L}_{NMM}$ is defined in Eq. \eqref{NMMF} with 
$\mathcal{F}=B^2/2$.

Then the energy density $\rho$ and pressure $p$ of the $NMM$ is obtained as follows:
\bea
&&\rho= {\frac {{B}^{2}}{2\,{B}^{2}\beta+2}},\label{rho1}
\eea 
\bea
p=\,{\frac {-3\,{B}^{4}\beta+{B}^{2}}{6 \left( {B}^{2}\beta+1 \right) 
^{2}}}
.\label{p1}
\eea 
Assuming that $f_{1}(R)=R$ and $f_{2}(R)=1+\lambda R$, the field equations \eqref{FEq} are simplified as
\bea
&& R_{\mu\nu}-\frac{1}{2} R g_{\mu\nu}=\kappa^2 \Big[-2\lambda \mathcal{L}_{NMM} R_{\mu\nu}+\left(1+\lambda R\right) T_{\mu\nu}+\nonumber\\
&&2 \lambda\left(\nabla_{\mu}\nabla_{\nu}-g_{\mu\nu}\Box\right)\mathcal{L}_{NMM}\Big].\nonumber\\
 \label{FEq1}
\eea 
We introduce the flat Friedmann-Robertson-Walker (FRW) metric
\begin{equation}
ds^2=-dt^2+a^2(t)\,\delta_{ij} dx^i dx^j \,,
\label{FlatFRW}
\end{equation}
where $a(t)$ is the scale factor. By using this metric \eqref{FlatFRW} in the field equations \eqref{FEq1}, we obtain the following modified Friedmann equations
\bea
\label{00}
&& \frac{3 {H}^{2}}{\kappa^2}=\frac{B\Big[ 12\lambda H\dot{B} +B \left(1+\beta {{B}^{2}}\right) \left(1+6\lambda {{H}^{2}}\right)\Big]}{2{{\left( \beta{{B}^{2}}+1\right)}^{2}}},\\
&& \dot{H}=\Big[\frac{1}{3}{{B}^{2}}\left( 6\beta \left(\kappa^2\lambda-\beta\right){{\dot{B} }^{2}}+12\kappa^2\lambda {{H}^{2}}+\kappa^2\right)-\nonumber\\
&& \kappa^2\lambda B \left(\ddot{B}-H \dot{B}\right)-\left(\kappa^2\lambda+2\beta\right){{\dot{B}}^{2}}\Big]\times \\
&& \Big[\beta \left(\kappa^2\lambda-\beta\right){{B}^{4}}-\left(\kappa^2\lambda+2\beta\right){{B}^{2}}-1\Big]^{-1}-\frac{2\beta {{\dot{B}}^{2}}}{\beta {{B}^{2}}+1}. \nonumber
\label{ii}
\eea 
where the Hubble constant is defined as $H=\frac{\dot{a}}{a}$.

On other hand, from the energy conservation law in Eq. \eqref{EClaw}, and by using the flat FRW metric \eqref{FlatFRW}, we find the usual continuity equation 
\be
\dot{\rho}+3 \left(\rho+p\right)H=0.
\label{CEq}
\ee Furthermore, substituting the expressions for energy density $\rho$ and pressure density $p$ from Eq.s \eqref{rho} and \eqref{p} we obtain the motion equation
\be
\dot{B}=-2 B H.
\label{B}
\ee Clearly, the equations \eqref{00}, \eqref{ii} and \eqref{B} are not all independent equations. From equation \eqref{00} and \eqref{ii} we also can obtain Eq.\eqref{B}. The source of the inflation is the strong magnetic fields in the early universe.
By using the energy density $\rho$ and pressure $p$ in Eq. (\ref{rho1}) and Eq. (\ref{p1}), and integrating the Eq. (\ref{CEq}), the evolution of the magnetic field $B$ with respect to the scale factor $a$ is obtained:
$B(t)=B_{0}/a(t)^{2}$. It is noted that $B_{0}$ stands for $a(t)=1$, where the inflation causes the increasing of the scale factor $a$, on the other hand the magnetic field decreases. 







\section{Perfect fluid description and dynamics of inflation}

\subsection{Slow-roll parameters}

Following Ref.\cite{Bamba:2014wda}, the field equations \eqref{00}, \eqref{ii} and \eqref{B} can be rewritten in the form of an effective perfect fluid as follows:

\begin{eqnarray}
&& \frac{3}{\kappa^2} H(N)^2 =\rho_{eff}(N),\\
&& -\frac{2}{\kappa^2} H(N) \frac{d H(N)}{d N}=\rho_{eff}(N)+P_{eff}(N),
\label{FrEqs}
\end{eqnarray} 

where the cosmic time derivatives have been expressed as derivatives with respect 
to the number of $e$-folds through $dN=Hdt$. In this way, we have that 
$N=\log(a(t)/a_{i})$. This effective fluid description includes the energy densities and pressures coming from the non-liner magnetic monopole field and the non-minimal coupling term. Here the effective energy density and the effective pressure densities take the following form:
\begin{eqnarray}
\label{rhoeff}
&& \rho_{eff}(N)=\frac{X+1}{2\left[\beta(X+1)^2+\kappa^2\lambda \left(3  X-1\right)\right]},\\
&& P_{eff}(N)=\left[\frac{1}{6 \left(\beta  (X+1)^2+\kappa^2\lambda  (3 X-1)\right)^2}\right]\times \nonumber\\
&& \left[\beta (X-3) (X+1)^2+\kappa^2\lambda  \left(X (10-9 X)+3\right)\right],
\label{peff}
\end{eqnarray} 
where:
\begin{eqnarray}
\label{Xvar}
&& X(N)=\exp(4 N)/(\beta B_{0}^2).
\end{eqnarray}
The equation of state (EoS) parameter of the total fluid is given by
\begin{equation}
P_{eff}(N)=-\rho_{eff}(N)+f(\rho_{eff}),
\label{EoS1}
\end{equation} where the function $f(\rho_{eff})$ is given by:
\begin{eqnarray}
&& f(\rho_{eff})=-\frac{1}{12 \kappa ^2 \lambda }-\frac{\rho_{eff}}{3}  (16 \beta  \rho_{eff}-3)-\nonumber\\
&& 3 \kappa ^2 \lambda  \rho_{eff}^2\pm \left[\frac{1}{12 \kappa ^2 \lambda}+\frac{5 \rho_{eff}}{6} \right] \times\nonumber\\
&& \Bigg[4 \kappa ^2 \lambda  \rho_{eff} \left(\rho_{eff} \left(16 \beta +9 \kappa ^2 \lambda \right)-3\right)+1\Bigg]^{\frac{1}{2}}=\nonumber\\
&& f(N)=\frac{2 X \left[4\kappa^2\lambda +\beta  (X+1)^2\right]}{3 \left[\beta  (X+1)^2+\kappa^2\lambda  (3 X-1)\right]^2}.
\label{frho}
\end{eqnarray}
Therefore, by using Eqs. \eqref{EoS1}, \eqref{rhoeff}, and \eqref{frho}, the effective EoS parameter reads as:
\begin{eqnarray}
\label{weff}
&& w_{eff}(N)\equiv \frac{P_{eff}(N)}{\rho_{eff}(N)}=-1+\frac{f(\rho_{eff})}{\rho_{eff}(N)}=\nonumber\\
&& \frac{(X-3)(X+1)^2+\kappa ^2 \alpha  (X (10-9 X)+3)}{3   (X+1)^3+3 \kappa ^2 \alpha  (3 X-1) (X+1)},
\end{eqnarray} 
where $\alpha\equiv\frac{\lambda}{\beta}$. During inflation
$\left|f(\rho_{eff})/\rho_{eff}(N)\right|\ll 1$ and thus $w_{eff}(N)\approx -1$. In fact, from the above equation, we have that at early times, when $X\rightarrow 0$, the EoS parameter goes to $-1$, while for late-times, when $X\rightarrow \infty$, the EoS tends to $\frac{1}{3}$. Then, in the effective fluid description, the nonlinear magnetic monopole behaves as an effective cosmological constant at the very early universe, driving inflation and while for later times, it behaves as a radiation fluid. This novel feature implies that a separate reheating phase is avoided in comparison to scalar field inflation.



In order to study the dynamics of inflation, we introduce the slow-parameter $\varepsilon\equiv -\frac{1}{H(N)}\frac{d H(N)}{d N}=-\frac{3}{2} f(\rho_{eff})/\rho_{eff}$, which satisfies the condition $\varepsilon\ll 1$ during slow-roll inflation. By using Eq. \eqref{FrEqs}, $\epsilon$ has the following dependence in $X$:
\begin{equation}
\varepsilon=\frac{2 X \left( X^{2}+2 X+4\alpha+1\right) }{\left(X+1\right)\left( X^{2}+\left(3\alpha +2\right) X-\alpha+1\right)}.
\label{slowparameter}
\end{equation}  
It is interesting to mention that a very early times, when $X\rightarrow 0$, $\varepsilon$ goes to zero, while in the limit $X\rightarrow \infty$, we have that $\varepsilon$ becomes 2, which is the exact
value for $\varepsilon$ in a radiation-dominated universe, since the scale factor evolves as $a(t)\propto t^{1/2}$. 

On the other hand, the end of inflation takes place when the condition $\varepsilon=1$ is satisfied, such that, $X(t_{\textup{end}})=X_{\textup{end}}$.
Thus, from Eq.\eqref{slowparameter} we obtain:
\begin{equation}
X_{\textup{end}}= \kappa ^2 \alpha-\frac{1 }{3} +\frac{4}{3 U}-\frac{8   \kappa ^2 \alpha }{U}+\frac{3 \kappa ^4 \alpha ^2}{U}+\frac{U}{3},
\label{Xend}
\end{equation}
where:
\begin{eqnarray}
&& U=\left[8 +36  \kappa ^2 \alpha -108 \kappa ^4 \alpha ^2+27 \kappa ^6 \alpha ^3+6 \sqrt{3} S\right]^{1/3},\\
&& S=\left[\kappa ^2 \alpha  \left(16 -72 \kappa ^2 \alpha +108  \kappa ^4 \alpha^2 - 27 \kappa ^6 \alpha ^3\right)\right]^{1/2}.
\end{eqnarray} For a sake of comparison, in the minimally coupled case, i.e, $\lambda=0$ (or equivalently $\alpha=0$) we find that $S=0$, $U=2/3$ and thus we have that $X_{\textup{end}}=1$.

\subsection{Comparison with current observations}

\begin{figure}[tbp]
\centering
\includegraphics[width=0.8\textwidth]{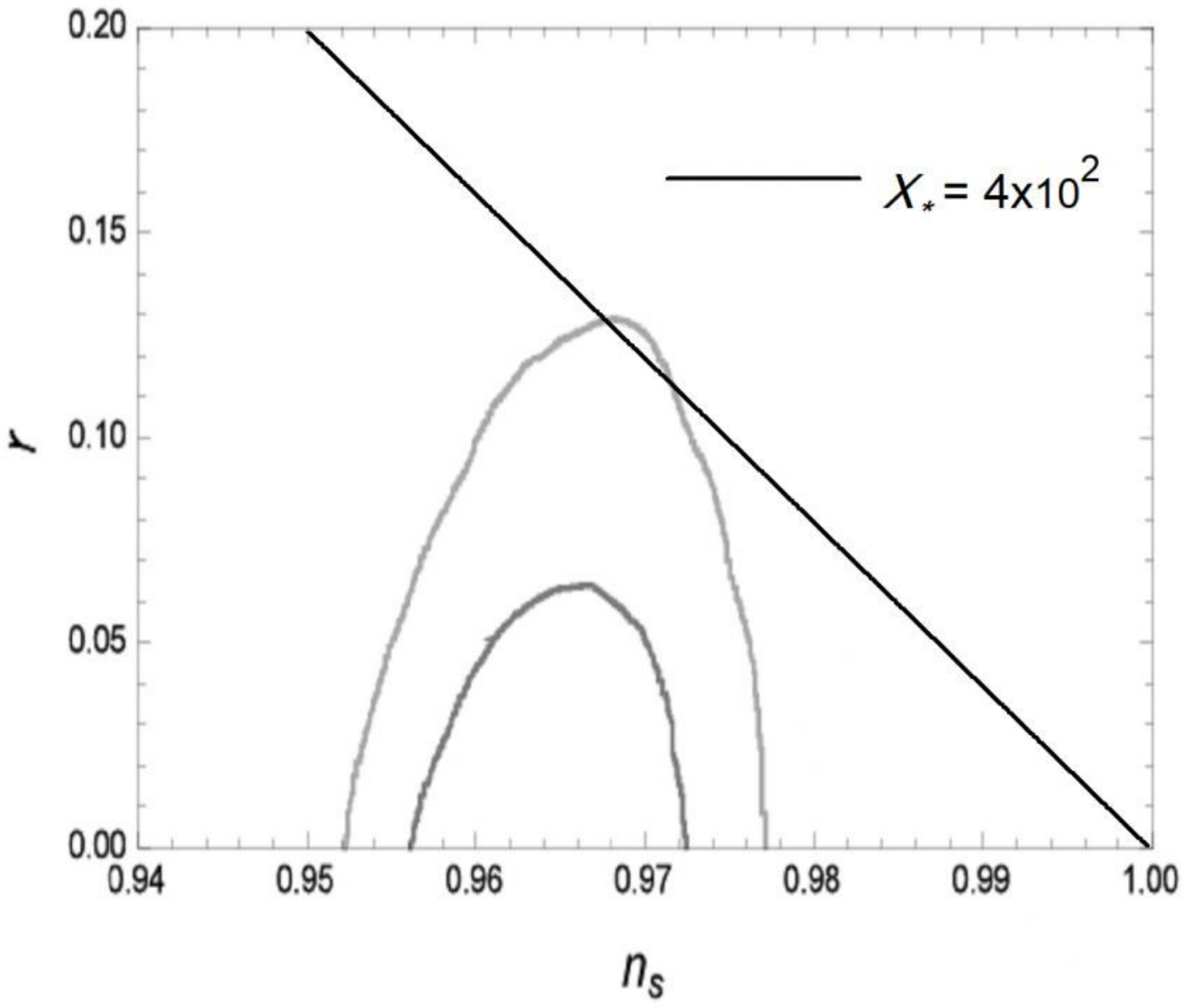}
\vspace{-0.5 in}
\caption{Allowed contours at the 68 and 95 $\%$ C.L., from the latest Planck data and theoretical predictions in the plane $r$ versus $n_s$ for the fluid description (\ref{observables}), by fixing $X_{*}=4\times 10^2$ and $\beta=1\times 10^{-2} M_{pl}^{-4}$.}
\label{FIG1}
\end{figure}

\begin{figure}[tbp]
\centering
\includegraphics[width=0.7\textwidth]{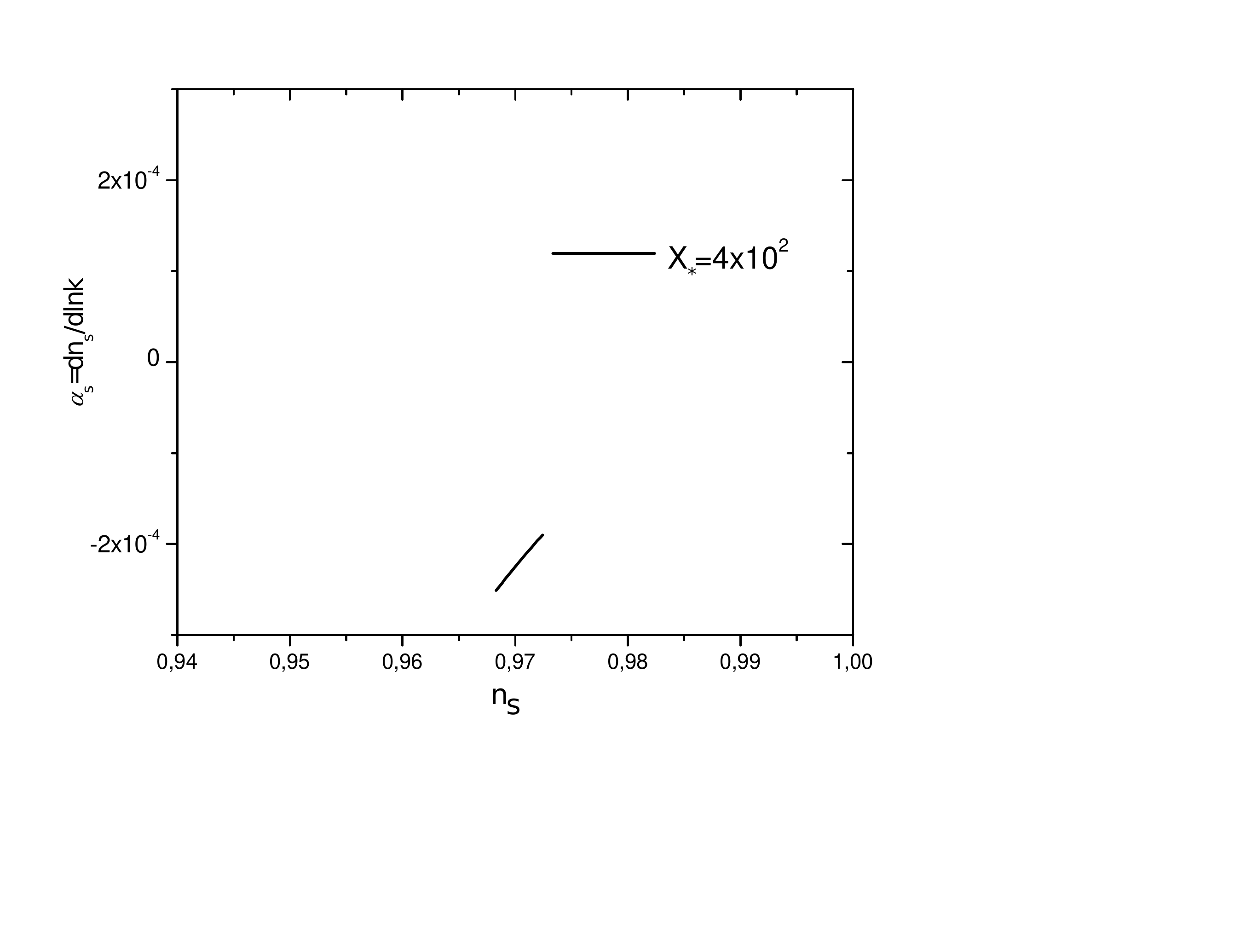}
\vspace{-0.5 in}
\caption{Plot of the running of the scalar spectral index $d n_s/d\ln k$
for the allowed range for $\alpha$, by fixing $X_{*}=4\times 10^2$ and $\beta=1\times 10^{-2} M_{pl}^{-4}$.}
\label{FIG2}
\end{figure}

As a first approach, the perturbative analysis of our model can be also developed in the framework of an effective fluid description \cite{Bamba:2014wda}, where the scalar spectral index, the tensor-to-scalar ratio, and the running of the spectral index take the form :
\begin{equation}
\label{observables}
n_{s}\simeq 1-6 \frac{f(\rho_{eff})}{\rho_{eff}(N)},\:\:\:\: r\approx 24 \frac{f(\rho_{eff})}{\rho_{eff}(N)},\:\:\:\: \alpha_{s}\approx -9 \left(\frac{f(\rho_{eff})}{\rho_{eff}(N)}\right)^2,
\end{equation}
where
\begin{equation}
\frac{f(\rho_{eff})}{\rho_{eff}(N)}=\frac{4X\left[(1+X)^2+4\kappa^2 \alpha\right]}{(1+X)\left[(1+X)^2+\kappa^2 \alpha (3X-1)\right]}.
\end{equation}
The form of Eq.(\ref{observables}) is valid when the condition $\left|f(\rho_{eff})/\rho_{eff}(N)\right|\ll 1$  is satisfied during inflation.
Then, $r$, $n_s$, and $\alpha_s$ for a given value of the $X$ variable are certain functions of $\beta$ and $\alpha$. First, and having in mind that the radiation-dominated epoch after inflation takes place for very large values of $X$, we fix $X_{*}=4\times10^2$, and consider the case $\beta=1\times 10^{-2} M_{pl}^{-4}$, being $M_{pl}$ the reduced Planck mass. For this fixing of parameters we plot $r$ versus $n_s$ in the same plot with the allowed contour at the 68 and 95 $\%$ C.L. from the latest Planck data, as is shown in Fig.\ref{FIG1}. The theoretical prediction lies inside the 95 $\%$ C.L. region when $\alpha$ takes values in the following range:
\begin{eqnarray}
\label{range_alpha}
1\textup{.}98\times 10^{5} M_{pl}^2 < \alpha <7\textup{.}74\times 10^{5} M_{pl}^2,
\label{alpha}
\end{eqnarray}  implying that the scalar spectral index lies inside the range:
\begin{eqnarray}
\label{range_ns}
0\textup{.}968<n_s<0\textup{.}972.
\end{eqnarray}
On the other hand, since $\alpha=\lambda/\beta$, then the constraint for the coupling $\lambda$ becomes:
\begin{eqnarray}
\label{range_lambda}
1\textup{.}98\times 10^{3} M_{pl}^{-2}<\lambda<7\textup{.}74\times 10^{3} M_{pl}^{-2}.
\end{eqnarray}
In order to determine the prediction of this model regarding the running of the spectral
index $\alpha_s$, Fig.\ref{FIG2} shows the trajectories in the $n_s-d n_s/d\ln k$ plane for the allowed range for $\alpha$ already obtained, and we found that these values are close to $-2\times 10^{-4}$, being in agreement with current bound imposed by last data of Planck. 

\begin{figure}[tbp]
\centering
\includegraphics[width=0.8\textwidth]{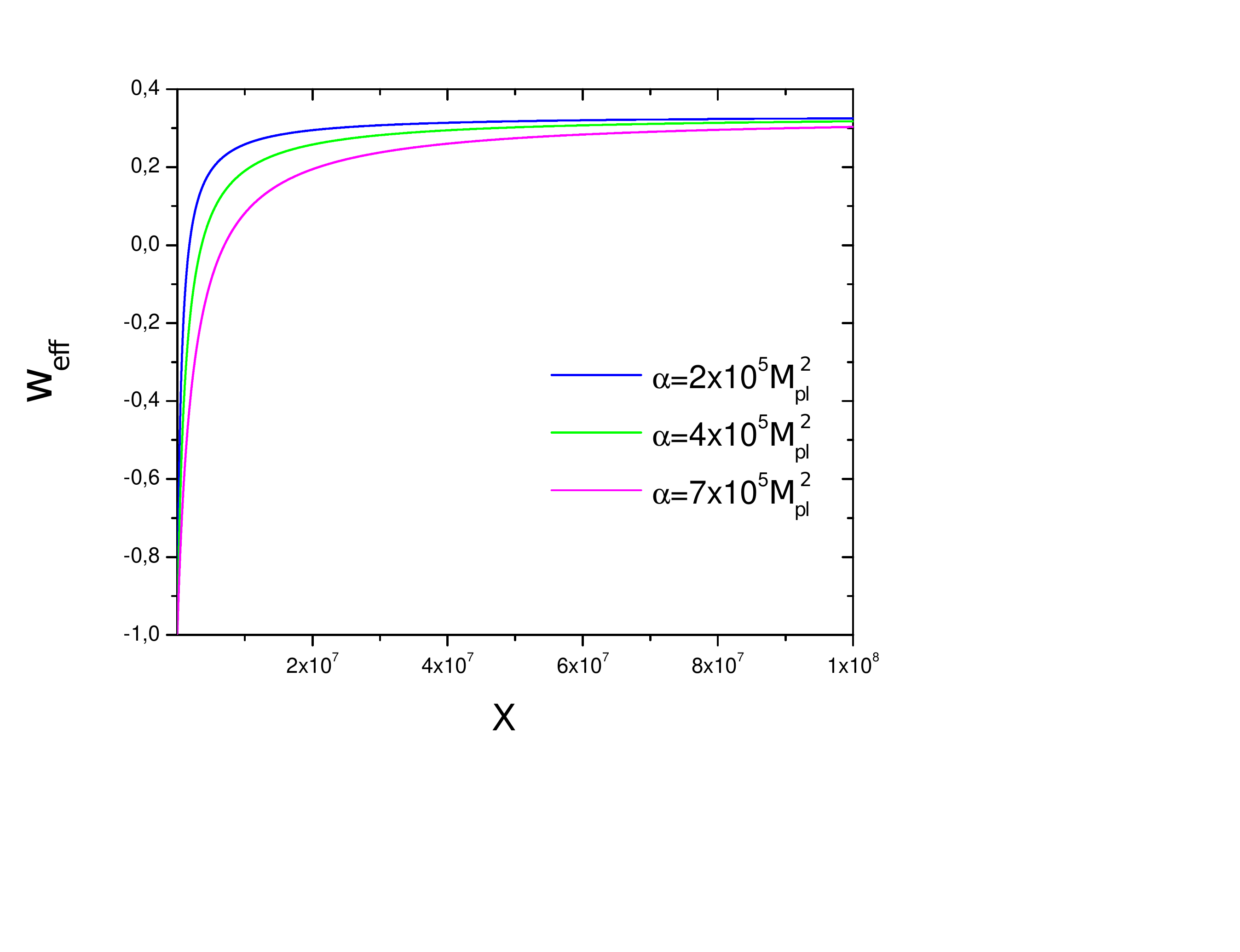}
\vspace{-0.5 in}
\caption{Plot of the effective EoS parameter $w_{eff}$ against the $X$ variable for three different values of $\alpha$ by fixing $\beta=1\times 10^{-2} M_{pl}^{-4}$.}
\label{FIG3}
\end{figure}

\begin{figure}[tbp]
\centering
\includegraphics[width=0.8\textwidth]{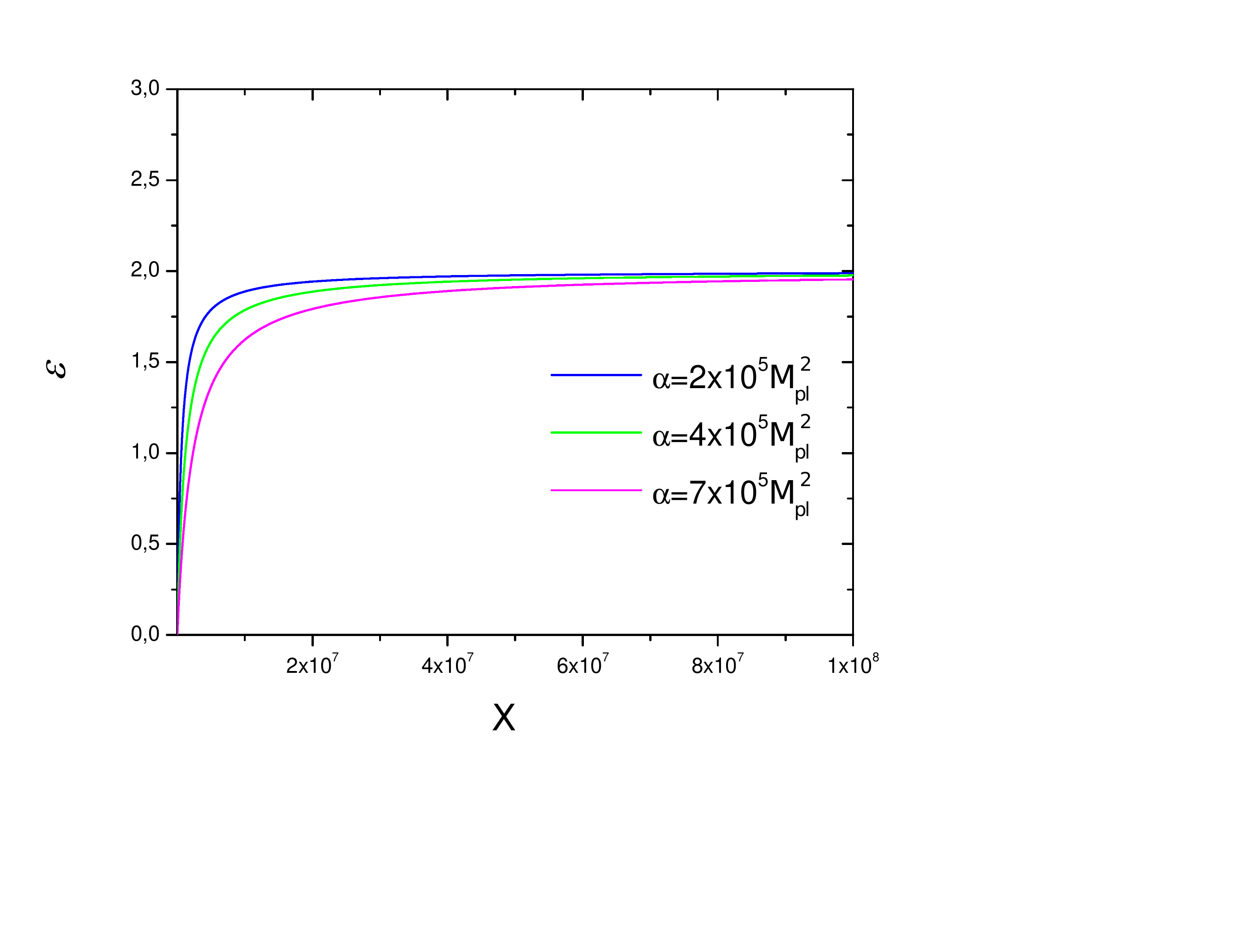}
\vspace{-0.5 in}
\caption{Plot of the slow-roll parameter $\varepsilon$ against the $X$ variable for three different values of $\alpha$ by fixing $\beta=1\times 10^{-2} M_{pl}^{-4}$.}
\label{FIG4}
\end{figure}

The smooth transition from inflation up to the radiation-dominated epoch at background level is shown in Fig.\ref{FIG3} and Fig.\ref{FIG4}. In particular, Fig.\ref{FIG3} depicts the evolution of the effective EoS parameter $w_{eff}$, given by Eq.(\ref{weff}), as function of $X$, for three different values of $\alpha$ within the allowed range found previously. In this way, it is demonstrated that the model interpolates between and effective cosmological constant driving inflation ($w_{eff}\simeq-1$) and a radiation fluid ($w_{eff}=1/3$). On the other hand, Fig.\ref{FIG4} shows the evolution of the slow-roll parameter $\varepsilon$, given by Eq.(\ref{slowparameter}), which is also plotted against the $X$ variable for three different values of $\alpha$. We can see again that, at early times, an slow-roll inflationary epoch takes place ($\varepsilon \ll 1$) and, for late times, the evolution corresponds to a radiation-dominated universe ($\varepsilon=2$).

As we have seen at perturbative level, the effective fluid description yields that our model becomes only marginally consistent with current observations. However, a more rigorous treatment of perturbations at linear model may be regarded as a further research. 

In next subsection, we will use the current bound on the propagation speed of gravitational waves to put constraints on both parameters $\alpha$ and $\lambda$.

\subsection{Complying with the gravitational waves constraints}

By linearising the field equations around Minkowski background $g_{\mu\nu}=\eta_{\mu\nu}+h_{\mu\nu}$, with $h_{\mu\nu}\ll 1$, and for a perfect fluid matter source which has a Lagrangian density that satisfies $\mathcal{L}_{m}=-\rho\approx const$, one is led to the following wave equation for the tensor modes \cite{Bertolami:2017svl}
\be
\Box \left(h_{\mu\nu}-\frac{1}{4} \eta_{\mu\nu} h\right)=m_{g}^2 h_{\mu\nu},
\ee where
\be
m_{g}^2=\frac{f_{1}(R_{0})-2 f_{2}(R_{0})\rho}{F_{1}(R_{0})-2 F_{2}(R_{0}) \rho},
\label{mg2}
\ee is the squared mass of the graviton, $m_{g}<7.7 \times 10^{-23} eV$ \cite{Abbott:2017vtc}, and $R_{0}=0$ for the Minkowski background. The group velocity of the gravitational wave is calculated from the corresponding dispersion relation which gives $v^2_{g}= 1-m_{g}^2/2 k^2$ \cite{Will:1997bb}. Here, an extra longitudinal mode related to $\delta \rho$ is also present into the theory \cite{Bertolami:2017svl}. In the case of $f_{1}(R)=R$ and $f_{2}(R)=1+\lambda R$, one can see that
\begin{equation}
m_{g}^2=-\frac{2 \rho}{1-2\lambda \rho}
\label{mg2.2}.
\end{equation}

We have focused on a nonlinear monopole magnetic field as matter source, whose Lagrangian density is given by \eqref{NMMF}. Since the solution found for the magnetic field scales as $B(t)=B_{0}/a(t)^2$, and $\mathcal{F}=B(t)^2/2=B_{0}^2/2 a(t)^{4}$, thus, for small values of the scale factor, $a(t)\rightarrow 0$, one obtains $\mathcal{L}_{NMM}=-\rho \approx -1/(2 \beta)$ being that $\beta$ is a constant. Therefore, by substituting this expression for energy density $\rho$ into Eq. \eqref{mg2.2}, we find $m_{g}^2=-\beta^{-1}/(1-\alpha)$, and hence 
\begin{equation}
\alpha> 2.84\times 10^{-1} M_{pl}^2,
\label{GWConstraint}
\end{equation} for $\beta=1\times 10^{-2} M_{pl}^{-4}$.

If we compare Eq. \eqref{GWConstraint} with the constraint in Eq \eqref{alpha}, obtained from constraints on the inflationary observables, one can see that the large values required for the parameter $\alpha$ and $\lambda$ guarantee a very small mass of the graviton in agreement with observations \cite{Abbott:2017vtc}, and therefore, also a propagation speed of gravitational waves very close to the speed of light with a high precision \cite{TheLIGOScientific:2017qsa,Monitor:2017mdv}.

In the opposed limit, when the scale factor tends to larger values, $a(t)\rightarrow \infty$, the effective coupling term between curvature and matter behaves as $a^{-6}$, and hence, the matter sector, which comes to behave as radiation, decouples from curvature, and trivially satisfying all the gravitational waves constraints of GR.

\section{Concluding remarks} \label{Conclu}
It has been studied the inflationary dynamics of a nonlinear monopole magnetic field which is nonminimally coupled to curvature. In the context of nonminimally coupled $f(R)$ gravity theories, the gravitational sector is taken to have the Einstein-Hilbert form, whereas that the matter Lagrangian density, that is, the nonlinear monopole magnetic field, is linearly coupled to curvature. We have used an effective fluid description in order to calculate the effective EOS and the slow-roll parameters. In this framework, the nonlinear magnetic monopole field behaves as an effective cosmological constant at the very early Universe, driving inflation and while for later times, it behaves as a radiation fluid. This novel feature implies that a separate reheating phase is avoided in comparison to scalar field inflation.

After, by using the most recent observational data from Planck collaboration, we set the limits on the parameter of the nonminimal coupling, and the quotient of the nonminimal coupling and the nonlinear monopole magnetic scales. First, we have used the theoretical prediction that we have found coherent with the 95 $\%$ C.L. region as shown in Fig. 1 for the value of the $\alpha$:
\begin{eqnarray}
1\textup{.}98\times 10^{5} M_{pl}^2 < \alpha <7\textup{.}74\times 10^{5} M_{pl}^2,
\end{eqnarray} 

and then the constraint for the coupling $\lambda$ becomes:
\begin{eqnarray}
1\textup{.}98\times 10^{3} M_{pl}^{-2}<\lambda<7\textup{.}74\times 10^{3} M_{pl}^{-2}.
\end{eqnarray}

Then we have checked our model by running the spectral
index $\alpha_s$, in Fig.\ref{FIG2} to show the trajectories in the $n_s-d n_s/d\ln k$ plane for the allowed range for $\alpha$, hence these values are close to $-2\times 10^{-4}$, being in agreement with current bound imposed by last data of Planck. 

We have seen that at perturbative level, the effective fluid description yields that our model becomes only marginally consistent with current observations. However, a more rigorous treatment of perturbations at linear model can be regarded as a further research.

Second, we have obtained the constraints for the nonminimal coupling parameter $\alpha$ using the current bound on the propagation speed of gravitational waves to as follows:
\begin{equation}
\alpha> 2.84\times 10^{-1} M_{pl}^2,
\end{equation} for $\beta=1\times 10^{-2} M_{pl}^{-4}$.
By comparing Eqs. \eqref{alpha} and \eqref{GWConstraint}, we have concluded that the large values required for the parameter $\alpha$, and $\lambda$, obtained from constraints on inflationary observables, are consistent with a very small mass of the graviton in agreement with observations \cite{Abbott:2017vtc}, and therefore, also verifying a propagation speed of gravitational waves very close to the speed of light with a high precision in agreement with the recent gravitational wave data GW$170817$ of LIGO and Virgo \cite{TheLIGOScientific:2017qsa,Monitor:2017mdv}.

Linde \cite{Linde:1994hy} and  Vilenkin \cite{Vilenkin:1994pv} discussed the inflation could occur in the cores of topological defects, where the scalar field is forced to stay near the maximum of its potential. However in our paper, we have used nonlinear magnetic monopole field nonminimally coupled to curvature as source of cosmic inflation. From conceptual point of view, the class of models treated in our work, has the advantage that it involves only electromagnetic fields, and it does not presuppose the existence of exotic entities such as scalar fields or another speculative modified matter models. It is worth noting that, the only fundamental scalar field which has been observed in nature is the Higgs field, but we do not have certainty if it can be identified as the inflaton once that it alone cannot generate inflation, being required a nonminimal coupling to gravity  and a Higgs mass $m_{H}$ greater than $230$ $Gev$ which is outside the widely accepted upper bound $m_{H}\sim 180$ $Gev$  \cite{Bezrukov:2007ep,Barvinsky:2008ia}.


\begin{acknowledgments}
G. O. acknowledges DI-VRIEA for financial support through Proyecto Postdoctorado $2017$ VRIEA-PUCV. This work was supported by Comisi\'on Nacional
de Ciencias y Tecnolog\'ia of Chile through FONDECYT Grant N$^{\textup{o}}$ 3170035 (A. \"{O}.), N$^{\textup{o}}$ 1170279 (J. S.) and N$^{\textup{o}}$ 1170162 (N. V.).  A. \"{O}. is grateful to Prof. Douglas Singleton for hosting him at the California State University, Fresno and also thanks to Prof. Leonard Susskind and Stanford Institute for Theoretical Physics for hospitality.   
\end{acknowledgments}


\begin{thebibliography}{99}

\bibitem{Guth:1980zm} 
  A.~H.~Guth, The Inflationary Universe: A Possible Solution to the Horizon 
  and Flatness Problems,
  Phys.\ Rev.\ D {\bf 23}, 347 (1981).
	
\bibitem{Starobinsky} A. A. Starobinsky, A New Type of Isotropic Cosmological Models Without Singularity,
 Phys. Lett. B \textbf{91}, 99 (1980).
	
\bibitem{Linde:1981mu} 
  A.~D.~Linde, A New Inflationary Universe Scenario: A Possible Solution of 
  the Horizon, Flatness, Homogeneity, Isotropy and Primordial Monopole Problems,
  Phys.\ Lett.\ B {\bf 108}, 389 (1982).
	
\bibitem{Mukhanov:1990me} 
  V.~F.~Mukhanov, H.~A.~Feldman and R.~H.~Brandenberger, Theory of cosmological perturbations. Part 1. Classical perturbations. Part 2. Quantum theory of perturbations. Part 3. Extensions,
  Phys.\ Rept.\  {\bf 215}, 203 (1992).


\bibitem{Liddle_Lyth2000} A. R. Liddle, D. H. Lyth, Cosmological inflation and Large-Scale Structure, (Cambridge
University Press, Cambridge; New York, 2000).

\bibitem{MukhanovBook} V. F. Mukhanov, {\it Physical Foundations of Cosmology} (Cambridge Univ. Press, Cambridge, 2005).

\bibitem{Liddle:1999mq} 
  A.~R.~Liddle, An Introduction to cosmological inflation,
 astro-ph/9901124.




\bibitem{DeFelice:2010aj} 
  A.~De Felice and S.~Tsujikawa, $f(R)$ theories,
  Living Rev.\ Rel.\  {\bf 13}, 3 (2010) [arXiv:1002.4928].


\bibitem{Sotiriou:2008rp} 
  T.~P.~Sotiriou and V.~Faraoni, $f(R)$ Theories of Gravity,
  Rev.\ Mod.\ Phys.\  {\bf 82}, 451 (2010) [arXiv:0805.1726].

	
\bibitem{Nojiri:2010wj} 
  S.~Nojiri and S.~D.~Odintsov, Unified cosmic history in modified gravity: from $F(R)$ theory to Lorentz non-invariant models,
  Phys.\ Rept.\  {\bf 505}, 59 (2011) [arXiv:1011.0544].
	
\bibitem{Nojiri:2017ncd} 
  S.~Nojiri, S.~D.~Odintsov and V.~K.~Oikonomou, Modified Gravity Theories on a Nutshell: Inflation, Bounce and Late-time Evolution,
  Phys.\ Rept.\  {\bf 692}, 1 (2017) [arXiv:1705.11098].

	
\bibitem{Lobo:2008sg} 
  F.~S.~N.~Lobo, The Dark side of gravity: Modified theories of gravity,
  Dark Energy-Current Advances and Ideas, arXiv:0807.1640.

\bibitem{Clifton:2011jh} 
  T.~Clifton, P.~G.~Ferreira, A.~Padilla and C.~Skordis, Modified Gravity and Cosmology,
  Phys.\ Rept.\  {\bf 513}, 1 (2012) [arXiv:1106.2476].
  
	
\bibitem{Capozziello:2011et} 
 S.~Capozziello and M.~De Laurentis, Extended Theories of Gravity,
 Phys.\ Rept.\  {\bf 509}, 167 (2011) [arXiv:1108.6266].

\bibitem{Copeland:2006wr} 
  E.~J.~Copeland, M.~Sami and S.~Tsujikawa, Dynamics of dark energy,
  Int.\ J.\ Mod.\ Phys.\ D {\bf 15}, 1753 (2006) [hep-th/0603057].
  
  
    

  
  


\bibitem{Nojiri:2004bi} 
  S.~Nojiri and S.~D.~Odintsov, Gravity assisted dark energy dominance and cosmic acceleration,
  Phys.\ Lett.\ B {\bf 599}, 137 (2004) [astro-ph/0403622].
  
\bibitem{Allemandi:2005qs} 
  G.~Allemandi, A.~Borowiec, M.~Francaviglia and S.~D.~Odintsov, Dark energy dominance and cosmic acceleration in first order formalism,
  Phys.\ Rev.\ D {\bf 72}, 063505 (2005) [gr-qc/0504057].

 
\bibitem{Nojiri:2006ri} 
  S.~Nojiri and S.~D.~Odintsov, Introduction to modified gravity and gravitational alternative for dark energy,
  eConf C {\bf 0602061}, 06 (2006)
  [Int.\ J.\ Geom.\ Meth.\ Mod.\ Phys.\  {\bf 4}, 115 (2007)] [hep-th/0601213].
 
  
\bibitem{Bertolami:2007gv} 
  O.~Bertolami, C.~G.~Boehmer, T.~Harko and F.~S.~N.~Lobo, Extra force in $f(R)$ modified theories of gravity,
  Phys.\ Rev.\ D {\bf 75}, 104016 (2007) [arXiv:0704.1733].
	
\bibitem{Harko:2008qz} 
  T.~Harko, Modified gravity with arbitrary coupling between matter and geometry,
  Phys.\ Lett.\ B {\bf 669}, 376 (2008) [arXiv:0810.0742]
	
\bibitem{Harko:2010mv} 
  T.~Harko and F.~S.~N.~Lobo, $f(R,L_{m})$ gravity,
  Eur.\ Phys.\ J.\ C {\bf 70}, 373 (2010) [arXiv:1008.4193].

\bibitem{Bertolami:2009ic} 
  O.~Bertolami and J.~P\'aramos, Mimicking dark matter through a non-minimal gravitational coupling with matter,
  JCAP {\bf 1003}, 009 (2010) [arXiv:0906.4757].
 
\bibitem{Bertolami:2013kca} 
  O.~Bertolami, P.~Fraz\~ao and J.~P\'aramos, Cosmological perturbations in theories with non-minimal coupling between curvature and matter,
  JCAP {\bf 1305}, 029 (2013) [arXiv:1303.3215].


  
\bibitem{Bertolami:2011fz} 
  O.~Bertolami and J.~P\'aramos, Mimicking the cosmological constant: Constant curvature spherical solutions in a non-minimally coupled model,
  Phys.\ Rev.\ D {\bf 84}, 064022 (2011) [arXiv:1107.0225].

\bibitem{Bertolami:2010cw} 
  O.~Bertolami, P.~Fraz\~ao and J.~P\'aramos, Accelerated expansion from a non-minimal gravitational coupling to matter,
  Phys.\ Rev.\ D {\bf 81}, 104046 (2010) [arXiv:1003.0850].

\bibitem{Gomes:2016cwj} 
  C.~Gomes, J.~G.~Rosa and O.~Bertolami, Inflation in non-minimal matter-curvature coupling theories,
  JCAP {\bf 1706}, 021 (2017) [arXiv:1611.02124].



\bibitem{TheLIGOScientific:2017qsa} 
  B.~P.~Abbott {\it et al.} [LIGO Scientific and Virgo Collaborations], GW$170817$: Observation of Gravitational Waves from a Binary Neutron Star Inspiral, Phys.\ Rev.\ Lett.\  {\bf 119}, 161101 (2017) [arXiv:1710.05832].
 
 
\bibitem{Lombriser:2015sxa} 
  L.~Lombriser and A.~Taylor, Breaking a Dark Degeneracy with Gravitational Waves,
  JCAP {\bf 1603}, 031 (2016) [arXiv:1509.08458].

\bibitem{Lombriser:2016yzn} 
  L.~Lombriser and N.~A.~Lima, Challenges to Self-Acceleration in Modified Gravity from Gravitational Waves and Large-Scale Structure,
  Phys.\ Lett.\ B {\bf 765}, 382 (2017) [arXiv:1602.07670].
 
 
\bibitem{Monitor:2017mdv} 
  B.~P.~Abbott {\it et al.} [LIGO Scientific and Virgo and Fermi-GBM and INTEGRAL Collaborations], Gravitational Waves and Gamma-rays from a Binary Neutron Star Merger: GW$170817$ and GRB $170817$A,
  Astrophys.\ J.\  {\bf 848}, L13 (2017) [arXiv:1710.05834].
 
 \bibitem{Baker:2017hug} 
  T.~Baker, E.~Bellini, P.~G.~Ferreira, M.~Lagos, J.~Noller and I.~Sawicki, Strong constraints on cosmological gravity from GW170817 and GRB 170817A,
  Phys.\ Rev.\ Lett.\  {\bf 119}, 251301 (2017) [arXiv:1710.06394].


\bibitem{Ezquiaga:2017ekz} 
  J.~M.~Ezquiaga and M.~Zumalac\'arregui, Dark Energy After GW170817: Dead Ends and the Road Ahead,
  Phys.\ Rev.\ Lett.\  {\bf 119}, 251304 (2017) [arXiv:1710.05901].
	
\bibitem{Creminelli:2017sry} 
  P.~Creminelli and F.~Vernizzi, Dark Energy after GW$170817$ and GRB$170817$A,
  Phys.\ Rev.\ Lett.\  {\bf 119}, 251302 (2017) [arXiv:1710.05877].
  
\bibitem{Sakstein:2017xjx} 
  J.~Sakstein and B.~Jain, Implications of the Neutron Star Merger GW$170817$ for Cosmological Scalar-Tensor Theories, Phys.\ Rev.\ Lett.\  {\bf 119}, 251303 (2017) [arXiv:1710.05893].

  
\bibitem{Bertolami:2017svl} 
  O.~Bertolami, C.~Gomes and F.~S.~N.~Lobo, Gravitational waves in theories with a non-minimal curvature-matter coupling, arXiv:1706.06826.
  
\bibitem{haw} 
  S.~W.~Hawking, Singularities in the universe,
  Phys.\ Rev.\ Lett.\  {\bf 17}, 444 (1966).
  
\bibitem{Carrol} S. M. Carroll, Living Rev.Rel. \textbf{4}, 1 (2001).


	\bibitem{Durrer:2013pga} 
  R.~Durrer and A.~Neronov, Cosmological Magnetic Fields: Their Generation, Evolution and Observation,
  Astron.\ Astrophys.\ Rev.\  {\bf 21}, 62 (2013) [arXiv:1303.7121].

\bibitem{Kunze:2013kza} 
  K.~E.~Kunze, Cosmological Magnetic Fields,
  Comments Plasma Phys.\ Contr.\ Fusion {\bf 55}, 124026 (2013) [arXiv:1307.2153].

\bibitem{Kunze:2007ph} 
  K.~E.~Kunze, Primordial magnetic fields and nonlinear electrodynamics,
  Phys.\ Rev.\ D {\bf 77}, 023530 (2008) [arXiv:0710.2435].

\bibitem{Novello:2003kh} 
  M.~Novello, S.~E.~Perez Bergliaffa and J.~Salim, Non-linear electrodynamics and the acceleration of the universe, Phys.\ Rev.\ D {\bf 69}, 127301 (2004) [astro-ph/0312093].


\bibitem{novello0} 
  V.~A.~De Lorenci, R.~Klippert, M.~Novello and J.~M.~Salim, Nonlinear electrodynamics and FRW cosmology,
  Phys.\ Rev.\ D {\bf 65}, 063501 (2002).
	
	\bibitem{Neronov:1900zz} 
  A.~Neronov and I.~Vovk, Evidence for strong extragalactic magnetic fields from Fermi observations of TeV blazars,
  Science {\bf 328}, 73 (2010) [arXiv:1006.3504].
 
\bibitem{Taylor:2011bn} 
  A.~M.~Taylor, I.~Vovk and A.~Neronov, Extragalactic magnetic fields constraints from simultaneous GeV-TeV observations of blazars,
  Astron.\ Astrophys.\  {\bf 529}, A144 (2011) [arXiv:1101.0932].
 
 
\bibitem{Subramanian:2015lua} 
  K.~Subramanian, The origin, evolution and signatures of primordial magnetic fields,
  Rept.\ Prog.\ Phys.\  {\bf 79},  076901 (2016) [arXiv:1504.02311].

\bibitem{Ade:2015cva} 
  P.~A.~R.~Ade {\it et al.} [Planck Collaboration], Planck 2015 results. XIX. Constraints on primordial magnetic fields,
  Astron.\ Astrophys.\  {\bf 594}, A19 (2016) [arXiv:1502.01594].
 
\bibitem{AlvesBatista:2016urk} 
  R.~Alves Batista, A.~Saveliev, G.~Sigl and T.~Vachaspati, Probing Intergalactic Magnetic Fields with Simulations of Electromagnetic Cascades,
  Phys.\ Rev.\ D {\bf 94}, 083005 (2016) [arXiv:1607.00320].

\bibitem{Grasso:2000wj} 
  D.~Grasso and H.~R.~Rubinstein, Magnetic fields in the early universe,
  Phys.\ Rept.\  {\bf 348}, 163 (2001) [astro-ph/0009061].
  
  \bibitem{tolman} R.~Tolman and P.~Ehrenfest, Temperature Equilibrium in a Static Gravitational Field,
  Phys.\ Rev.\  {\bf 36}, 1791 (1930).

  
  \bibitem{Bento:1992wy} 
  M.~C.~Bento, O.~Bertolami, P.~V.~Moniz, J.~M.~Mourao and P.~M.~Sa, On the cosmology of massive vector fields with SO(3) global symmetry,
  Class.\ Quant.\ Grav.\  {\bf 10}, 285 (1993)  [gr-qc/9302034].
  
   \bibitem{Bertolami:1991cf} 
  O.~Bertolami and J.~M.~Mourao, The Ground state wave function of a radiation dominated universe,
  Class.\ Quant.\ Grav.\  {\bf 8}, 1271 (1991).
  
  \bibitem{Bertolami:1990je} 
  O.~Bertolami, J.~M.~Mourao, R.~F.~Picken and I.~P.~Volobuev, Dynamics of euclidenized Einstein Yang-Mills systems with arbitrary gauge groups,
  Int.\ J.\ Mod.\ Phys.\ A {\bf 6}, 4149 (1991).
 
  \bibitem{Golovnev:2008cf} 
  A.~Golovnev, V.~Mukhanov and V.~Vanchurin, Vector Inflation,
  JCAP {\bf 0806}, 009 (2008) [arXiv:0802.2068].
  
  \bibitem{DeLorenci:2002mi} 
  V.~A.~De Lorenci, R.~Klippert, M.~Novello and J.~M.~Salim, Nonlinear electrodynamics and FRW cosmology,
  Phys.\ Rev.\ D {\bf 65}, 063501 (2002).


\bibitem{Novello:2006ng} 
  M.~Novello, E.~Goulart, J.~M.~Salim and S.~E.~Perez Bergliaffa, Cosmological Effects of Nonlinear Electrodynamics, Class.\ Quant.\ Grav.\  {\bf 24}, 3021 (2007) [gr-qc/0610043].

\bibitem{Ovgun:2016oit} 
  A.~\"{O}vg\"{u}n, Inflation and Acceleration of the Universe by Nonlinear Magnetic Monopole Fields,''
  Eur.\ Phys.\ J.\ C {\bf 77}, 105 (2017) [arXiv:1604.01837].
  
\bibitem{Ovgun:2017iwg} 
  A.~\"{O}vg\"{u}n, G.~Leon, J.~Magana and K.~Jusufi, Falsifying cosmological models based on a non-linear electrodynamics, [arXiv:1709.09794].
	
\bibitem{Kruglov:2017vca} 
  S.~I.~Kruglov, Inflation of universe due to nonlinear electrodynamics, Int.\ J.\ Mod.\ Phys.\ A {\bf 32}, 1750071 (2017) [arXiv:1705.01455].

\bibitem{Sharif:2017pdd} 
  M.~Sharif and S.~Mumtaz, Stability of the accelerated expansion in nonlinear electrodynamics,
  Eur.\ Phys.\ J.\ C {\bf 77}, 136 (2017) [arXiv:1702.04716].
	
\bibitem{Kruglov:2014hpa} 
  S.~I.~Kruglov, A model of nonlinear electrodynamics, Annals Phys.\  {\bf 353}, 299 (2014) [arXiv:1410.0351].
	
	
\bibitem{Kruglov:2016cdm} 
  S.~I.~Kruglov, Acceleration of Universe by Nonlinear Electromagnetic Fields,
  Int.\ J.\ Mod.\ Phys.\ D {\bf 25}, 1640002 (2016) [arXiv:1603.07326].

  
\bibitem{Kruglov:2015fbl} 
  S.~I.~Kruglov, Universe acceleration and nonlinear electrodynamics, Phys.\ Rev.\ D {\bf 92}, 123523 (2015) [arXiv:1601.06309].

\bibitem{Kruglov:2016lqd} 
  S.~I.~Kruglov, Nonlinear electromagnetic fields as a source of universe acceleration,
  Int.\ J.\ Mod.\ Phys.\ A {\bf 31}, 1650058 (2016) [arXiv:1607.03923].
  
  

	
\bibitem{Campanelli:2007cg} 
  L.~Campanelli, P.~Cea, G.~L.~Fogli and L.~Tedesco, Inflation-Produced Magnetic Fields in Nonlinear Electrodynamics, Phys.\ Rev.\ D {\bf 77}, 043001 (2008)  [arXiv:0710.2993].
	
\bibitem{Hollenstein:2008hp} 
  L.~Hollenstein and F.~S.~N.~Lobo, Exact solutions of f(R) gravity coupled to nonlinear electrodynamics,
  Phys.\ Rev.\ D {\bf 78}, 124007 (2008) [arXiv:0807.2325].
  
\bibitem{Bamba:2008ja} 
  K.~Bamba and S.~D.~Odintsov, Inflation and late-time cosmic acceleration in non-minimal Maxwell-$F(R)$ gravity and the generation of large-scale magnetic fields,
  JCAP {\bf 0804}, 024 (2008) [arXiv:0801.0954].
	




  
  
  	
  











\bibitem{Bamba:2014wda} 
  K.~Bamba, S.~Nojiri, S.~D.~Odintsov and D.~S\'aez-G\'omez,
  Inflationary universe from perfect fluid and $F(R)$ gravity and its comparison with observational data,
  Phys.\ Rev.\ D {\bf 90}, 124061 (2014) [arXiv:1410.3993].
 


\bibitem{Abbott:2017vtc} 
  B.~P.~Abbott {\it et al.} [LIGO Scientific and VIRGO Collaborations], GW$170104$: Observation of a $50$-Solar-Mass Binary Black Hole Coalescence at Redshift $0.2$, Phys.\ Rev.\ Lett.\  {\bf 118},  221101 (2017) [arXiv:1706.01812].
  
\bibitem{Will:1997bb} 
  C.~M.~Will, Bounding the mass of the graviton using gravitational wave observations of inspiralling compact binaries, Phys.\ Rev.\ D {\bf 57}, 2061 (1998) [gr-qc/9709011].
	


  \bibitem{Linde:1994hy} 
  A.~D.~Linde, Monopoles as big as a universe,
  Phys.\ Lett.\ B {\bf 327}, 208 (1994) [astro-ph/9402031].
  
  \bibitem{Vilenkin:1994pv} 
  A.~Vilenkin, Topological inflation,
  Phys.\ Rev.\ Lett.\  {\bf 72}, 3137 (1994) [hep-th/9402085].
 
 \bibitem{Bezrukov:2007ep} 
  F.~L.~Bezrukov and M.~Shaposhnikov, The Standard Model Higgs boson as the inflaton,
  Phys.\ Lett.\ B {\bf 659}, 703 (2008) [arXiv:0710.3755].

\bibitem{Barvinsky:2008ia} 
  A.~O.~Barvinsky, A.~Y.~Kamenshchik and A.~A.~Starobinsky, Inflation scenario via the Standard Model Higgs boson and LHC,
  JCAP {\bf 0811}, 021 (2008) [arXiv:0809.2104].



\end{thebibliography}
\end{document}